
\magnification=\magstep1
\hoffset=-0.15 truein
\hsize=6.55 truein
\vsize=9.3 truein
\openup 2\jot
\input mssymb
\font\bg=cmb10 scaled \magstep2
\font\biggish=cmb10 scaled \magstep1
\parindent=0pt
\hfill DAMTP 93-53

\hfill hep-th/9310068

\hfill October 1993
\vskip 50pt
{\centerline {\bg Fundamentals of Poisson Lie Groups}}
{\centerline {\bg with Application to the Classical Double}}
\vskip 20pt
{\centerline {K.S. Ahluwalia}}
\vskip 5pt
{\it {\centerline {Department of Applied Mathematics and Theoretical
Physics}}
{\centerline {University of Cambridge}}
{\centerline {Silver Street}}
{\centerline {Cambridge CB3 9EW}}
{\centerline {U.K.}}}
\vskip 30pt
{\centerline {\bf Abstract}}
\vskip 10pt
We give a constructive account of the fundamental ingredients of
Poisson Lie theory as the basis for a description of the classical
double group $D$.  The double of a group $G$ has a pointwise
decomposition $D\sim G\times G^*$, where $G$ and $G^*$ are Lie
subgroups generated by dual Lie algebras which form a Lie bialgebra.
The double is an example of a factorisable Poisson Lie group, in the
sense of Reshetikhin and Semenov-Tian-Shansky [1], and usually the
study of its Poisson structures is developed only in the case when the
subgroup $G$ is itself factorisable.  We give an explicit description
of the Poisson Lie structure of the double without invoking this
assumption.  This is achieved by a direct calculation, in
infinitesimal form, of the dressing actions of the subgroups on each
other, and provides a new and general derivation of the Poisson Lie
structure on the group $G^*$.  For the example of the double of SU(2),
the symplectic leaves of the Poisson Lie structures on SU(2) and
SU(2$)^*$ are displayed.
\vfill
\eject
{{\biggish I. Introduction }}
\vskip 5pt
A Poisson Lie group is a Lie group with a compatible Poisson
structure.  Here compatibility refers to group multiplication, which
is required to be a Poisson map from the product manifold $G\times G$
to the group $G$.  Consequently the Poisson structure is termed
multiplicative.  Much of the motivation for the study of such a
structure is provided by the dressing transformation group of a
classical integrable system.  This is the symmetry group of the system
in the Lax pair formalism and it carries a natural Poisson structure
which is multiplicative.  A quantised integrable system has a quantum
group symmetry and so Poisson Lie groups also generate interest as the
classical counterparts of quantum groups.
\vskip 3pt
Associated to any Poisson Lie group $G$ is a Poisson Lie group $G^*$
and also a Lie bialgebra.  The groups $G$ and $G^*$ form a dual
Poisson Lie pair, so termed because their Lie algebras, ${\cal G}$ and
${\cal G}^*$ respectively, are dual to each other as vector spaces.
The linearisation of the Poisson structure on one group produces the
Lie algebra structure of the dual.  The condition which fixes the
Poisson structures to be multiplicative also ensures that these dual
Lie algebras satisfy the defining property of a Lie bialgebra.  This
is because a multiplicative Poisson structure and a Lie bialgebra are
both defined in terms of a 1-cocycle.
\vskip 3pt
In many cases, and notably for semisimple Lie groups, the 1-cocycle is
constructed from a solution of a Yang-Baxter type equation, denoted
$r\in{\cal G}\otimes {\cal G}$ and called an r-matrix.  Particular
interest is shown when a Poisson Lie group is linked with a
quasitriangular r-matrix, that is, an r-matrix which is a
non-antisymmetric solution of the classical Yang-Baxter equation.
Such an r-matrix can be used to construct an isomorphism between the
universal enveloping algebras of the dual Lie algebras ${\cal G}$ and
${\cal G}^*$ in such a way as to pick out a unique factorisation
property of the group $G$.  Hence the group and the Lie bialgebra
associated to a quasitriangular r-matrix are both described as
factorisable.  An important class of factorisable groups is provided
generically by the double group.  This is constructed from any dual
Poisson Lie pair so as to have a pointwise factorisation into the
product of these groups.  An example is the group SL(2, ${\Bbb C}$)
which is the double of SU(2).  Any element of SL(2,~${\Bbb C}$)
factors uniquely into the product of an SU(2) and an SU(2$)^*$ group
element; here the factorisation property simply provides a restatement
of the Iwasawa decomposition.
\vskip 3pt
The double group provides a natural setting for the study of a dual
Poisson Lie pair, since it incorporates both groups together.  This is
reflected in much of the early literature.  Drinfel'd [2] was the
first to explicitly write down the structure of a Poisson Lie group,
relating it to solutions of the modified Yang-Baxter equation and also
to Lie bialgebras, from which he constructed the double Lie algebra.
Semenov-Tian-Shansky [3] established that the Poisson Lie structures
of $G$ and $G^*$ could be obtained by reduction from a symplectic
structure on the double group, and also gave the multiplicative
Poisson structure for $G^*$ in the case when $G$ is factorisable.
Importantly, this paper also presented dressing transformations as
Poisson Lie actions, and showed that their orbits give the symplectic
leaves of the structures on $G$ and $G^*$.  The term factorisable was
actually coined in a later work with Reshetikhin [1], where it was
used for Lie bialgebras but with the factorisation at group level
explained using the double as the principal example.  Lu and Weinstein
[4] noticed that the Iwasawa decomposition [5] is a concrete
realisation of factorisation of the double group, whilst in an
independently motivated approach Majid [6] derived this result.  In an
as yet unfinished line of work, Kosmann-Schwarzbach and Magri [7] have
put the double Lie algebra in the context of twilled extensions, as
well as relating the Yang-Baxter equations to conditions on a Schouten
curvature.  A worked example of the general formalism of a dual
Poisson Lie pair has been given by Babelon and Bernard [8] for a Toda
system.
\vskip 3pt
The Poisson Lie structure of the double group has been well documented
in these papers, yet much of the presentation has been at a formal
level only.  It is straightforward to write down the multiplicative
Poisson tensor since the r-matrix of the double is well known.
However, to actually write down the corresponding Poisson brackets
requires a calculation, at the infinitesimal level, of the actions of
the subgroups $G$ and $G^*$ on each other.  These actions are the
abstract description of dressing transformations.
Semenov-Tian-Shansky [9] briefly indicated how they can be inferred by
considering the group $G$ to be factorisable, and most explicit
presentations of the Poisson brackets of the double concentrate on
such a scenario.  This, however, restricts the Poisson Lie pairs that
are described.  In particular, the important example of the dual pair
constructed from the maximal compact subgroup of a complex semisimple
Lie group is not addressed.
\vskip 3pt
With this in mind, we provide an alternative approach to the
description of the double group without restricting to the class of
Poisson Lie dual pairs generated by a factorisable group.  We
construct a new isomorphism between the enveloping algebras of a dual
Poisson Lie pair which is more generally applicable than the
factorisation isomorphism.  This facilitates a direct calculation of
the dressing transformations required to display the structure of the
double group.  The result is a simple working form for the Poisson
structure of the double group, from Poisson tensor to Poisson
brackets, and a completely explicit general presentation of the
Poisson Lie structure on the group $G^*$.
\vskip 3pt
We give an introduction to Lie bialgebras and then construct specific
examples.  The double-sided nature of Lie bialgebras is made explicit.
We describe the usual r-matrix approach to the defining cocycle
condition which is used when constructing a Lie bialgebra from a
semisimple Lie algebra.  However, we also describe how to start with
the dual non-semisimple Lie algebra and satisfy the cocycle condition.
Within the parallel development of a complex semisimple Lie algebra
and its maximal compact real subalgebra we describe the factorisation
isomorphism of a Poisson Lie dual pair, applicable in the former case
only, and show that there is an alternative isomorphism which is valid
in both cases.  After a description of the double Lie algebra and
multiplicative Poisson structures this new isomorphism is applied to
the calculation of the Poisson Lie structure of the double group.  We
conclude with the example of the Poisson Lie structures on SU(2) and
SU(2$)^*$ worked through in coordinates, and display their symplectic
leaves.
\vskip 3pt
Our analysis looks only at the local properties of Poisson Lie groups
and does not address their global geometry.
\vskip 15pt
{{\biggish II. Lie Bialgebras}}
\vskip 5pt
A Lie bialgebra is constructed from a Lie algebra by endowing its dual
vector space with a Lie bracket.  This is done subject to a particular
compatibility condition, which is usually satisfied using the solution
of a Yang-Baxter type equation -- an r-matrix.  An alternative method,
applicable to the duals of non-semisimple Lie algebras, is given in the
next section.
\vskip 3pt
Consider a Lie algebra ${\cal G}$, its dual vector space ${\cal G^*}$,
and a pairing between them denoted by $<t^a,t_b>=\delta^a_b$ for basis
vectors $t^a\in {\cal G^*}$, $t_a\in {\cal G}$.  We can define a Lie
bracket $[\; ,\,]_*$ on ${\cal G^*}$ via a map, the Lie cobracket
$$\eqalign{\delta\, \colon {\cal G}\to \,\,&{\cal G}\wedge {\cal G}\;,
\cr <[\xi,\zeta]_*, X>\, =&<\xi\otimes\zeta,\delta (X)>\;,\qquad
\xi,\zeta\in{\cal G^*}\;,\quad X\in {\cal G}.\cr} $$
Here and elsewhere we use capital Roman letters $X,Y,Z\dots$ to denote
elements of
${\cal G}$ and Greek letters $\xi, \zeta, \chi \dots$ to denote elements
of ${\cal G^*}$.

The structure constants for ${\cal G^*}$ and ${\cal G}$ are $$[t^a,
t^b]_*=C^{ab}_c t^c\;,\hskip 20pt [t_a,t_b]=f_{ab}^ct_c\;.$$
The pair (${\cal G},\,{\cal G^*}$) constitutes a Lie bialgebra when
the Lie brackets satisfy a compatibility condition.  This condition
was introduced by Drinfel'd as the infinitesimal version of the
algebra map condition on a coproduct, and requires that the Lie
cobracket of ${\cal G^*}$  should be represented as a 1-cocycle on
${\cal G}$, with
respect to the adjoint representation.  That is
$$\delta ([X,Y])=X.\delta (Y) -
Y.\delta (X)\;,\eqno{ X,Y\in {\cal G}\;,
\hskip 20pt (1)}$$
where $X.(Y\wedge Z)=[X\otimes 1 + 1\otimes X, Y\wedge Z]$.  It is
important to note that this condition is symmetric between ${\cal G}$
and ${\cal G^*}$, as can be seen, for instance, by writing it in terms
of structure constants.  Hence it also ensures that the Lie cobracket
of ${\cal G}$ is a 1-cocycle on ${\cal G^*}$, and so (${\cal
G}^*,\,{\cal G}$) is also a Lie bialgebra.
\vskip 3pt
The easiest way to satisfy (1), and the only way for semisimple
${\cal G}$, is to fix $\delta $ to be the coboundary of an element $r
= r^{ab} t_a\otimes t_b\in {\cal G}\otimes {\cal G}$, considered as a
0-cochain, so that $$\eqalign{\delta (X)&=[X\otimes 1+1\otimes X,
r]\;,\hskip 84pt X\in {\cal G}\;,\cr <[\xi,\zeta]_*,
X>&=\,<\xi\otimes\zeta, [X\otimes 1+1\otimes X, r]>\;,\hskip 30pt
\xi,\zeta\in {\cal G^*}.\cr}
\eqno(2)$$
\vskip 3pt
Antisymmetry of the Lie cobracket requires $$[X\otimes 1+1\otimes X,\,
r + \tau(r)] =0\;,\eqno(3)$$ where $\tau$ is the permutation
operator, $\tau(X\otimes Y)= Y\otimes X$, so that only the
antisymmetric part of $r$ contributes to the structure of ${\cal
G^*}$.

The Jacobi identity is satisfied provided $$[X\otimes 1\otimes 1+
1\otimes X\otimes 1+1\otimes 1\otimes X,\, B_r] =0\;, \eqno(4)$$
where $$B_r=[r_{12}, r_{13}] + [r_{12}, r_{23}] + [r_{13},r_{23}] \in
{\cal G}\otimes {\cal G}\otimes {\cal G}\;,\eqno(5)$$ in which, for
example, $r^{ab}t_a\otimes 1\otimes t_b$ is denoted by $r_{13}$.  The
ad-invariance of $B_r$, condition (4), is known as the modified
Yang-Baxter equation (MYBE) for $r$.  The more restrictive condition
$B_r=0$ is
called the classical Yang-Baxter equation (CYBE).

If we decompose $r$ into symmetric
and antisymmetric parts, $r= s + a$, $\tau(s)=s$, $\tau(a)=-a$, then
we can use (3) to write (4) as a requirement on the antisymmetric part
of $r$ only.  From (3) it follows that
$B_s$ is ad-invariant, and that $B_r= B_s + B_a$, so that $B_r$ is
ad-invariant if and only if $B_a$ is ad-invariant.
\vskip 3pt
The structure constants of ${\cal G^*}$ are
$$C^{ab}_c = f^b_{cd}r^{ad} + f^a_{cd}r^{db}.\eqno(6)$$
Again it should be recognised that the symmetric part of $r$ does not
contribute to this expression.  For semisimple ${\cal G}$ expression
(6) leads to a non-semisimple dual Lie algebra.

In section III we indicate how, under certain quite broad conditions,
this equation can be inverted to give the structure constants of
${\cal G}$ in terms of those on ${\cal G^*}$ - see (8).
\vskip 3pt
We can consider $r$ to be a linear operator $$\eqalign{\hskip 90pt&
r\,\colon\, {\cal G^*}\to \,{\cal G}\;,\cr\hskip 70pt &r(\xi)=\,
<\xi\otimes {\rm id}, r>\;, \hskip 80pt \xi\in {\cal G^*}\;,\cr}$$
with transpose $$\eqalign{\hskip 90pt &r^*\,\colon\, {\cal G^*}\to
\,{\cal G}\;,\cr &r^*(\xi)=\, <\xi\otimes {\rm id}, \tau(r)>=\, <{\rm
id}\otimes
\xi, r>.
\hskip 34pt\cr}$$
This allows us to write the bracket on ${\cal G^*}$ as $$[\xi,
\zeta]_* = -{\rm ad}^*r(\xi).\zeta - {\rm ad}^*r^*(\zeta).\xi\;,$$
where we use the convention $<{\rm ad}^*X.\xi, Y> =
<\xi, {\rm ad}X.Y> = <\xi, [X, Y]>$.
\vskip 15pt
{{\biggish III. Group Factorisation}}
\vskip 5pt
Following the discussion in section II we see that it is possible to
specify a Lie bialgebra in terms of a solution of the MYBE; the
structure constants of ${\cal G^*}$ are given by (6).  In this section
we shall implement this for two generic examples; a complex,
semisimple Lie algebra ${\cal G}$, and its real, maximal, compact
subalgebra ${\cal K}$.  This incorporates the useful cases of
sl(2,${\Bbb C}$) and su(2).  The practical distinction between these
two examples is that ${\cal G}\otimes {\cal G}$ admits a
non-antisymmetric solution of the classical Yang-Baxter equation,
ie.$\;$a quasitriangular r-matrix, but ${\cal K}\otimes {\cal K}$ does
not.  In both instances the second rank Casimir element $t_a\otimes
t_a$, which defines a natural vector space isomorphism between ${\cal
G}^*$ (${\cal K}^*$) and ${\cal G}$ (${\cal K}$), can be defined to be
an isomorphism of Lie algebra structures, equipping the vector space
${\cal G}$ (${\cal K}$) with a new Lie bracket.  We describe the
extension of this Lie algebra isomorphism to the enveloping algebras,
as given by Reshetikhin and Semenov-Tian-Shansky [3], which produces a
unique factorisation of the group associated to ${\cal G}$.  We also
give a new extension of the same Lie algebra isomorphism in a way
which is valid for both ${\cal G}$ and ${\cal K}$.  This alternative
shall prove useful when discussing the Poisson structure of the double
group.
\vskip 3pt
To facilitate a straightforward comparison of the analysis for ${\cal
K}$ and ${\cal G}$ we shall consider expressions in the basis
$$\eqalign{t_a&=\, iH^r,\; V^{\alpha}\;, W^{\alpha}\cr
V^{\alpha}= {i\over \sqrt 2}
(E^{\alpha}+&E^{-{\alpha}})\;,\quad W^{\alpha}={1\over \sqrt 2}
(E^{\alpha}-E^{-{\alpha}})\cr}\eqno{(\alpha >0)\hskip 40pt}$$
where $[H^r, E^{\alpha}]={\alpha}^rE^{\alpha}$,
$[E^{\alpha},E^{-{\alpha}}]={\alpha}^rH^r$.  This basis is suitable
for both ${\cal K}$, over ${\Bbb R}$, and ${\cal G}$, over ${\Bbb C}$.

Choosing Trace$(t_at_b)=y\delta_{ab}$, $y$ dependent upon the
representation, we work with a non-degenerate, invariant inner product
denoted $(\,,\, )=y^{-1}$Trace$(\,.\,)$.  Taken with the square of
the longest root to be unity this is intrinsic to ${\cal G}$ and so
independent of any choice of basis or representation [10].
\vskip 3pt
A general solution of the MYBE for ${\cal G}$ takes the form [11]
$$r=\lambda\sum_{\alpha >0} (V^{\alpha}\otimes W^{\alpha} -
W^{\alpha}\otimes V^{\alpha}) + \mu\, t_a\otimes t_a\;.$$ If $\lambda
= \pm i \mu$ we have a solution of the CYBE, but to be admissable as
an r-matrix for ${\cal K}$ we must have $\lambda$ and $\mu$ both real.
This leads us to choose the following r-matrices for ${\cal K}$ and
${\cal G}$ respectively $$\eqalign{k_+&= {\textstyle {1\over 2}} (I +
r_0)\in {\cal K}\otimes {\cal K}\;,\cr r_+&= {\textstyle {1\over
2}}(-iI + r_0)\in {\cal G} \otimes {\cal G}\;,\cr}$$ where $r_0
=\sum_{\alpha >0}(V^{\alpha}\otimes W^{\alpha} - W^{\alpha}\otimes
V^{\alpha})$ and $I= t_a\otimes t_a$, the second rank Casimir.  It
should be noted that the complex nature of ${\cal G}$ means that
$ir_+$ is also an acceptable r-matrix.

We emphasise that
$$B_{r_\pm}=0\;,\qquad B_{k_\pm}={\textstyle {1\over
2}} B_{r_0}={\textstyle {1\over 2}} B_I={\textstyle {1\over
2}}f^c_{ab}t_a\otimes t_b\otimes t_c\;,\eqno(7)$$
where we denote
$\tau(r_+)= -r_-$ and $\tau(k_+)= -k_-$.  In particular, the
conditions imposed on r-matrices in [9] are satisfied by $ir_+$ for
${\cal G}$, but are not satisfied by $k_+$ for ${\cal K}$.
\vskip 3pt
We remark that if the antisymmetric part of an r-matrix satisfies
the MYBE in the fashion given by (7), then it is possible to invert the
relation between the structure constants of ${\cal G^*}$ and ${\cal G}$,
given in (6).  That is, we can construct the 1-cocycle on
${\cal G^*}$ which corresponds to the usual Lie bracket on ${\cal G}$.
It is given by $$\displaylines{<\xi,
[X,Y]>=<\bar\delta(\xi),X\otimes Y>\;,\cr
\bar\delta(\xi)=(<[t^a,t^b]_*, r_0(\xi)> + <[t^b,\xi]_*,r_0(t^a)> +
<[\xi, t^a]_*, r_0(t^b)>)\, t^a\otimes t^b\;,\cr}$$ so that
$$f_{ab}^c= C^{ab}_dr_0^{cd} + C^{bc}_dr_0^{ad} +
C^{ca}_dr_0^{bd}\;.\eqno(8)$$
This allows for a double-sided understanding of the structure of the
Lie bialgebra in this case.

The Lie cobrackets for ${\cal K^*}$ and ${\cal G^*}$ are
$$\eqalign{\delta_{\cal K} (X)&=[X\otimes 1+1\otimes X,
k_{\pm}]={\textstyle {1\over 2}}[X\otimes1+1\otimes X,r_0]\;,\cr
\delta_{\cal G} (X)&=[X\otimes 1+1\otimes X, r_{\pm}]={\textstyle {1\over 2}}
[X\otimes1+1\otimes X,r_0]\;.\cr}$$ These are the same, by design, so
that ${\cal G^*}$ is the complexification of ${\cal K^*}$, as with
their dual Lie algebras.
\vskip 3pt
We may regard $I=t_a\otimes t_a$ as defining the
vector space isomorphism $$\eqalign{I&\colon\, {\cal G^*}\,\to \,{\cal
G}\;,\cr I&\colon\, t^a\,\mapsto \, <t^a\otimes id, I> =t_a\;.\cr}$$ Upon
restriction to ${\cal K^*}$, that is, over $\Bbb R$, we also clearly
have $I\,\colon\,{\cal
K^*}\to \, {\cal K}$.

In addition, $$(X,Y)= <I^{-1}(X), Y> = <I^{-1}(Y), X>\;.$$ This allows
us to consider $I$ to be a Lie algebra isomorphism, defining a second
Lie algebra structure, denoted by $[\,,\,]_R$, on the vector space
${\cal G}$ via $$<[I^{-1}(X), I^{-1}(Y)]_*, Z> = <I^{-1}[X,Y]_R, Z> =
([X,Y]_R,Z)\;.$$ Using (6) we can write this second Lie bracket as
$$[X,Y]_R = [R_\pm X,Y]+[X,R_\mp Y]={\textstyle {1\over
2}}\left([R_0X,Y]+[X,R_0Y]\right)\;,\eqno(9)$$ where we set
$R_{\pm}\equiv r_{\pm}\circ I^{-1}$.  Again, if we replace $r_{\pm}$
by $k_{\pm}$ this will reduce to the same form, so it is equally valid
as a new bracket on the vector space ${\cal K}$.

We signify with a
subscript $R$ a vector space with this new Lie bracket, so that
$$I\,\colon\,{\cal G^*}\to\, {\cal G}_R$$
is a Lie algebra isomorphism, and over the real field $I\,\colon\,{\cal
K^*}\to \, {\cal K}_R$.

Since $I=i(r_+-r_-)=k_+-k_-$, any element of ${\cal G}$ or ${\cal K}$
admits a unique decomposition $$\eqalign{{\cal G}\ni\, &X=I(\xi)=iX_+
- iX_-\;,\qquad X_{\pm}=r_{\pm}(\xi)\;,\cr {\cal K}\ni\,
&Y=I(\zeta)=Y_+-Y_-\;,\qquad\quad Y_{\pm}=k_{\pm}(\zeta)\;.\cr}$$
However, there is a further property unique to $r_{\pm}$ as solutions
of the CYBE.  The expression (5) for $B_r$ can be rewritten
$$<I^{-1}(X)\otimes I^{-1}(Y)\otimes {\rm id}, B_r> = [RX,RY] -
R([X,RY] - [R^TX,Y])\;.$$
Hence, it follows that $$\eqalign{[R_\pm X,R_\pm Y]&=
R_\pm([X,Y]_R)\;,\cr
[r_\pm(\xi),r_\pm(\zeta)]&=r_\pm([\xi,\zeta]_*)\;.\cr}\eqno(10)$$ That
is, $r_{\pm}$ are Lie algebra homomorphisms.  In these circumstances
the Lie bialgebra (${\cal G},\,{\cal G^*}$) is said to be
factorisable, terminology which will become clearer when we
investigate the associated group.
\vskip 3pt
Thus, turning our attention to the corresponding universal enveloping
algebras, we have natural extensions of $r_{\pm}$ as algebra maps,
$$\eqalign{\tilde r_\pm &\colon\, U({\cal G^*})\to \, U({\cal
G})\;,\cr
\tilde r_\pm&(1_*)= 1\;,\cr \tilde
r_\pm&(\xi\zeta)=r_\pm(\xi)r_\pm(\zeta)\;,\quad {\rm etc.}\cr
}\eqno(11)$$ The map $I'=-iI$, which is equivalent to the isomorphism
$I$ when acting on a complex Lie algebra, can be extended [3] in the
form $$\tilde I'=m\circ ({\rm id}\otimes S)\circ (\tilde r_+\otimes
\tilde r_-)\circ \Delta \,\colon\,U({\cal G}^*)\to\, U({\cal
G}_R)\;,$$ where $\Delta$ is the usual coproduct, $\Delta (\xi)=
\xi\otimes 1 + 1\otimes\xi$, and $S$ the usual antipode, $S(\xi)=
-\xi$, extended as algebra and anti-algebra maps respectively.
Denoting the product in $U({\cal G}_R)$ by a $*$, this produces $$X*Y
= R_+X.Y - Y.R_-X = {\textstyle 1\over 2}(R_0X.Y - Y.R_0X -i XY -iYX
)\eqno(12)$$
Applied at the level of the group, within a suitable completion of the
enveloping algebra, this indicates that group elements have a unique
factorisation
$$\eqalign{\tilde I'\,\colon \,G^*&\to\, G_R\;,\cr
e^{\xi}\,&\mapsto\,e^{\tilde I'(\xi)}
=\exp\{r_+(\xi)\}.\exp\{-r_-(\xi)\}=g_+g_-^{-1}=g\;.\cr}\eqno(13)$$
This factorisation of group elements is valid in a region around the
identity.  For a study of the conditions required for this
factorisation to be a global property see Lu's thesis [12].

The group $G_R$ has the same elements as $G$ but a different product
law, given by
$$g*h = g_+h_+(g_-h_-)^{-1}=g_+hg_-^{-1}\;.\eqno(14)$$
The map $\tilde I'$ cannot be constructed from $k_{\pm}$, because they
are not Lie algebra isomorphisms, and so it is necessarily complex, as
seen in (12) for example.  Hence it does not map $U({\cal K}^*)$ into
the real algebra $U({\cal K}_R)$.  However, this extension of $I$ as
an algebra map is not unique, the only requirement is that $U({\cal
G}_R)$ be defined such that $\tilde I(\xi\zeta) - \tilde I(\zeta\xi)
=[I(\xi), I(\zeta)]_R $.  It shall prove useful when discussing the
Poisson structure of the double group to have an extension which does
map real algebras to real algebras.  This can be achieved in a
straightforward manner by defining
$$\eqalign{\tilde I&\,\colon \,U({\cal G}^*)\to\, U({\cal G}_R)\;,\cr
\tilde I&(1_*)=1_R\;,\cr\tilde I&(\xi\zeta\chi\dots)= i(\tilde
r_+-\tilde r_-)(\xi\zeta\chi\dots)\;.\cr}\eqno(15)$$
This leads to
$$\eqalign{\tilde I(\xi\zeta)& =i\tilde r_+(\xi\zeta) - i\tilde
r_-(\xi\zeta)\;,\cr &=i\tilde r_+(\xi)\tilde r_+(\zeta)-i\tilde
r_-(\xi)\tilde r_-(\zeta)\;,\cr &={\textstyle {1\over 2}}\{
I(\xi)r_0(\zeta) + r_0(\xi)I(\zeta)\}\;,\cr &=\tilde I(\xi)*\tilde
I(\zeta)\;,\cr}\eqno(16)$$
so that, in this instance, the product in $U({\cal G}_R)$ is
$$X*Y = {\textstyle 1\over 2}(X.R_0Y + R_0X.Y  )\eqno(17)$$
As required, $X*Y-Y*X=[X,Y]_R$, but now the product in $U({\cal G}_R)$
is real so that it is equally valid for $U({\cal K}_R)$, but {\it
without} replacing $r_{\pm}$ with $k_{\pm}$ at any stage.  At the
group level we find that
$$\tilde I\,\colon\,e^{\xi} \mapsto\, e_R^{\tilde I(\xi)} = 1 +
i[\exp\bigl(r_+(\xi)\bigr) - \exp\bigl(r_-(\xi)\bigr)]= 1+ i(g_+ -g_-)
= g\;.$$
The product law is equivalent to that in the previous case, that is
$$g*h = 1 + i(g_+h_+ - g_-h_-)\;.$$
\vskip 15pt
{\biggish IV. The Double}
\vskip 5pt
The double of a Lie algebra can be used to construct a factorisable Lie
bialgebra and is an object of considerable interest.  We give a
description of its properties as a prelude to examining the Poisson
Lie structure of the associated group.
\vskip 3pt
To motivate the construction of the double we observe that, with the
decomposition
of vectors $X=X_+-X_-$  the Lie bracket (9) on ${\cal
G}_R$  can be
written
$$[X,Y]_R= [X_+,Y_+]-[X_-, Y_-]\;.$$
This bracket resembles the direct sum of two Lie algebras.  Writing
${\cal G}={\cal G}_++{\cal G}_-$ as a direct sum of vector spaces then
${\cal G}_\pm$ are Lie subalgebras if $[X_\pm,
Y_\pm]\in {\cal G}_\pm$ for all $X_\pm,\,Y_\pm\in{\cal G}_\pm$.  In
this case ${\cal G}_R={\cal G}_+\oplus{\cal G}_{-_{opp}}$ as a direct
sum of Lie
algebras, where ${\cal G}_{-_{opp}}$ signifies ${\cal G}_-$ with the
opposite Lie bracket.
  Also, if the Lie bialgebra (${\cal G},\,{\cal G^*}$)  is
factorisable, $\pm R_\pm$ are
projection operators from ${\cal G}_R$ on to ${\cal G}_\pm$, since by (10)
$$[X_\pm,Y_\pm]=[R_\pm X,R_\pm Y]=R_\pm([X,Y]_R)=R_\pm([X_+,Y_+]-[X_-,
Y_-])$$
\vskip 3pt
The double ${\cal D}$ is constructed so that ${\cal G}_+,\,{\cal G}_-$
are dual Lie algebras. That is, the double is a Lie algebra containing
${\cal G}$ and ${\cal G}^*$ as Lie subalgebras, $${\cal D = G +
G^*}\;,$$ and ${\cal G}$ and ${\cal G}^*$ form a Lie bialgebra (${\cal
G},\,{\cal G^*}$).
It has a natural inner product $$\qquad <(X, \xi), (Y, \zeta)> = \zeta
(X) + \xi (Y)\;,\qquad\qquad X,Y\in {\cal G}\;,\quad \xi, \zeta\in
{\cal G}^*\;,$$ which, when required to be ad-invariant, fixes the
``mixed'' commutator to be $$\eqalign{[(X,0),(0,\xi)]\equiv [X, \xi]
&= ad^*\xi.X - ad^*X.\xi\;,\cr [t_a, t^b]&= C_a^{bc}t_c -
f^b_{ac}t^c\;.\cr}\eqno(18)$$ This can be derived straightforwardly by
considering $[X, \xi]$ contracted separately with $(Y,0)$ and
$(0,\zeta)$.  The Jacobi identity is satisfied precisely because of
the Lie bialgebra compatibility condition (1).
\vskip 3pt
We note that ${\cal G}$, ${\cal G}^*$ are isotropic subalgebras with
respect to this inner product, ie.
$$\eqalign{<(X,0),(X,0)> = 0 \hskip 20pt \forall \; (X,0)\in {\cal
G\subset D}\;,\cr
<(0, \xi), (0, \xi)> = 0 \hskip 20pt \forall \; (0, \xi)\in {\cal
G^*\subset D}\;.\cr}$$
Clearly, requiring ${\cal G}_\pm$ to be isotropic is equivalent to
requiring them to be dual, and consequently there is a one-to-one
correspondence between Manin triples and the double of a Lie algebra
(see for example [4]).
\vskip 3pt
By our opening argument the
r-matrices for ${\cal D}$, which we distinguish by a bar from
those for ${\cal G}$, are just projection operators
$$\eqalign{&{\bar r_+} = (0, t^a)\otimes (t_a,0)\;,\qquad\quad {\bar
r_-} = -\tau({\bar r_+}) = -(t_a,0)\otimes (0, t^a)\;,\cr & {\bar
r_+}\,\colon\,{\cal D^*}\to {\cal G\subset D}\;,\qquad {\bar
r_+}\,\colon\,(\xi, X)\mapsto \,<(\xi, X)\otimes id, (0, t^a)\otimes
(t_a,0)> = (X, 0)\;,\cr &{\bar r_-}\,\colon\,{\cal D^*}\to {\cal
G^*\subset D}\;,\qquad{\bar r_-}\,\colon\,(\xi, X)\mapsto -<(\xi,
X)\otimes id, (t_a,0)\otimes (0, t^a)> = -(0, \xi)\;,\cr}\eqno(19)$$
where $(\xi, X)\in {\cal D^*}$.  We do not explicitly distinguish
the pairing between ${\cal D}$ and ${\cal D^*}$ from the inner
product on ${\cal D}$, it should be clear in the usage which one is
required.

It is straightforward to confirm that ${\bar r_\pm}$ satisfy the CYBE,
as well as condition (3), hence the pair (${\cal D},{\cal D}^*$) is
indeed a factorisable Lie bialgebra.  The Lie algebra isomorphism
$I_{\cal D}=({\bar r_+ - \bar r_-})$ acts in the natural manner
$$\eqalign{I_{\cal D} &\colon\, {\cal D^* \to\, D}_R\;,\cr
I_{\cal D} &\colon\, (\xi, X)\mapsto\,(X, \xi)\;,\cr}$$
where, by design,
$$\eqalign{{\cal D}^* &= {\cal G}^*_{opp}\oplus {\cal G}\;,\cr
{\cal D}_R &= {\cal G}\oplus {\cal G}^*_{opp}\;,\cr}\eqno(20)$$
as the direct sum of Lie algebras.
\vskip 3pt
If the double is constructed from a Lie bialgebra (${\cal G},\,{\cal
G}^*$) which is itself factorisable then we can go further in its
description.  Clearly we can map
$${\cal D = G + G^*\to\, G + G_R}\;,\qquad\qquad (X, \xi)\mapsto (X,
I(\xi))\;,$$
where $I=(r_+ - r_-)\colon{\cal G^*}\to{\cal G}_R$, and $r_\pm$ are
the r-matrices for (${\cal G},\,{\cal G}^*$).  However, we also
have the linear isomorphism
$$\eqalign{J&\colon\, {\cal D \to \, D'=G\oplus G}\;,\cr J&\colon\, (X,
\xi)\mapsto (X, X) + (r_+\xi, r_-\xi)\;.\cr}$$
The algebras ${\cal G}$ and ${\cal G}_R$ ($\equiv {\cal G}^*$) are
embedded in ${\cal D'}$ by $$\eqalign{{\cal G}& \hookrightarrow {\cal
D'}\;,\hskip 30pt X\mapsto (X,X)\;,\cr {\cal G}_R&\hookrightarrow
{\cal D'}\;,\hskip 30pt Y=I(\xi)\mapsto (R_+Y, R_-Y) = (r_+\xi,
r_-\xi)\;.\cr}$$ Since any element of ${\cal D}'$ can be written as a
unique sum of elements in ${\cal G}$ and ${\cal G}_R$, we have $${\cal
D' = G + G}_R = {\cal G}\oplus {\cal G}\;.\eqno(21)$$ The inner
product on ${\cal D}'$ is $$<(X,Y), (X', Y')>= {\rm tr}(XX') - {\rm
tr}(YY')\;,$$ which is obviously ad-invariant, and also ensures that
the subalgebras ${\cal G}$ and ${\cal G}_R$ are isotropic.  The
r-matrices for ${\cal D}'$ are just the images of those in ${\cal D}$,
eg.  $${\bar r}'_+ = (J\otimes J){\bar r}_+ = \sum_a
(r_+t^a,r_-t^a)\otimes (t_a, t_a)\;.$$ They project on to ${\cal
G\subset D}'$ and ${\cal G}_R\subset {\cal D}'$; $$\eqalign{{\bar
r}'_+&\colon\,{\cal D}'^*\to\,{\cal G\subset D}'\;,\hskip 30pt {\bar
r}'_+\,\colon\,J^{*-1}(\xi, X)\mapsto\,(X,X)\;,\cr {\bar
r}'_-&\colon\,{\cal D}'^*\to\,{\cal G}_R\subset {\cal D}'\;,\hskip
30pt {\bar r}'_-\,\colon\,J^{*-1}(\xi, X)\mapsto\,(r_+\xi,
r_-\xi)\;,\cr}$$
where we identify an element of ${\cal D}'^*$ as
$$\eqalign{J^{*-1}&\colon\,{\cal D}^*\to\,{\cal D}'^*\;,\cr
J^{*-1}&\colon\,(\xi, X)\mapsto\,(I^{-1}X,I^{-1}X) + (I^{-1}r_+\xi,
I^{-1}r_-\xi)\;.\cr}$$
To elaborate on the working here we develop
$$\eqalign{&<\!J^{*-1}(\xi, X)\otimes id, {\bar r}'_+\!>
=<\!(I^{-1}X,I^{-1}X) + (I^{-1}r_+\xi, I^{-1}r_-\xi),
(r_+t^a,r_-t^a)\!>\!(t_a, t_a)\hfill\cr &=
\{<I^{-1}X,r_+t_a>-<I^{-1}X,r_-t_a>\}(t_a, t_a)=<I^{-1}X,t_a>(t_a,
t_a)=(X,X)\;,\cr}$$ which agrees with above.  Similarly it can be
confirmed that $$\eqalign{{\cal D}'^*&= {\cal G}^*_{opp}\oplus {\cal
G}^*_R\cr {\cal D}'_R&= {\cal G}\oplus {\cal G}_{R_{opp}}\cr}$$
\vskip 15pt
{\biggish V. Poisson Lie Groups}
\vskip 5pt
As a final precursor to investigating the Poisson structure of the
double group we give a general explanation of multiplicative Poisson
structures.  We make the connection with Lie bialgebras by identifying
a Lie bialgebra as the linearisation of a Poisson Lie structure at the
identity of the group.
\vskip 3pt
A Poisson manifold is a manifold $M$ with a Poisson bracket structure,
ie.$\;$a Lie algebra structure $\{\, ,\}_M$ on the vector space
$C^{\infty}(M)$, such that
$$\{f_1, f_2f_3\}_M = f_3\{f_1,f_2\}_M + f_2\{f_1,f_3\}_M\;,\hskip
20pt f_i\in
C^{\infty}(M)\;.$$
The product of two Poisson manifolds $(M,\{\, ,\}_M)$ and $(N,\{\,
,\}_N)$ is the Poisson manifold $(M\times N, \{\, ,\}_{M\times N})$
where $$\{f_1,f_2\}_{M\times N}(m,n) = \{f_1^m, f_2^m\}_N(n) +
\{f_1^n, f_2^n\}_M(m)\;,\eqno{f_i\in C^{\infty}(M\times N)\;,\hskip
35pt}$$ and $f_1^m$ denotes the smooth function on $N$ obtained from
$f_1\in C^{\infty}(M\times N)$ by keeping the $m\in M$ component
fixed, etc.

A Poisson morphism $\phi$ is a smooth map from a Poisson manifold
$(M,\{\, ,\}_M)$ to the Poisson manifold $(N,\{\, ,\}_N)$ which
satisfies $$\displaylines{\hfill \phi\,\colon\,M\to\,N\;,\hfill
(22)\cr \hskip 70pt \phi^*\{f_1, f_2\}_N = \{\phi^*f_1,
\phi^*f_2\}_M\;,\hskip 40pt f_i\in C^{\infty}(N)\;.\cr}$$ This is to
be understood as $\{f_1, f_2\}_N\circ\phi = \{f_1\circ
\phi, f_2\circ\phi\}_M$.

\vskip 3pt
We can now define a Poisson Lie group as a Lie group $G$ with a
Poisson structure $\{\, ,\}_G$ such that group multiplication is a
Poisson morphism from $(G\times G, \{\, ,\}_{G\times G})$, equipped
with the product structure, to $(G, \{\, ,\}_G)$.

Using $L_g, R_g$ to denote left and right multiplication by $g\in G$
we can write property (22) as
$$\{f_1,f_2\}_G(gh) = \{f_1\circ L_g, f_2\circ L_g\}_G(h) + \{f_1\circ
R_h, f_2\circ R_h\}_G(g)\;,\eqno(23)$$
for $g,h\in G$ and $f_i\in C^{\infty}(G)$.

In terms of a Poisson bivector $\Lambda\in \Gamma (TG\wedge TG)$
defined by $$\displaylines{ \Lambda(df_1, df_2) = \{f_1,f_2\}_G\;, \cr
{\rm ie.}\hskip 10pt <df_1(g)\otimes df_2(g),\Lambda(g)> =
\Lambda(df_1, df_2)(g) = \{f_1,f_2\}_G(g) \;,\cr}$$
the Poisson morphism condition becomes $$\Lambda(gh) = L_{g*}.\Lambda
(h) + R_{h*}.\Lambda (g) = [L_{g*}\Lambda ](gh) + [R_{h*}\Lambda
](gh)\;,$$ where $L_{g*},\,R_{g*}$ denote left and right translation
lifted to the tangent space.

Fixing $g=h=e$, the  identity element of $G$, we
find
$\Lambda (e) = \Lambda (e) + \Lambda (e).$  Thus $\Lambda (e) =0\;;$
and we see that the Poisson structure cannot be symplectic since it is
degenerate at the identity.
\vskip 3pt
The Poisson morphism condition can be satisfied by constructing the
Poisson tensor from solutions of the MYBE.  Consider
$$\{f_1,f_2\}(g)=<d^Lf_1(g)\otimes d^Lf_2(g), r> + <d^Rf_1(g)\otimes
d^Rf_2(g), r'>\eqno(24)$$
where $r,r'\in {\cal G}\otimes {\cal G}$, $\,{\cal G}$ being the Lie
algebra corresponding to $G$, and the left and right differentials are
given by $d^Lf(g)=L^*_{g}df(g)=d(f\circ L_g)(e)$ etc., where
$L^*_{g},\, R^*_{g}$ are the pullbacks of left and right translation.

In tensor notation (24) reads
$$\{g\otimes g\}= g\otimes g.r + r'.g\otimes g\;.\eqno(25)$$
Decomposing $r$ and $r'$ into symmetric and antisymmetric parts
$$\displaylines{ r = s + a\hskip 20pt r'= s' + a'
\cr \tau(s)=s\qquad \tau(s')=s'\qquad \tau (a)= -a \qquad \tau
(a')= -a'\cr}$$
then antisymmetry of the Poisson bracket requires that
$$g\otimes g.s + s'.g\otimes g=0\eqno(26)$$
so we see that only the antisymmetric parts of $r,r'$ contribute to
the Poisson structure.

Using (26) the Jacobi identity becomes the condition
$$\eqalign{\{\{g\otimes g\}\otimes g\} + {\rm cyclic}
=&\,(a_{12}a_{13} + a_{12}a_{23} + a_{23}a_{21} + a_{23}a_{31} +
a_{31}a_{32} + a_{31}a_{12})g\otimes g\otimes g \cr &+ g\otimes
g\otimes g(\!a'_{13}a'_{12} + a'_{23}a'_{12} + a'_{21}a'_{23} +
a'_{31}a'_{23} + a'_{32}a'_{31} + a'_{12}a'_{31}\!)\cr =&\,
B_a.g\otimes g\otimes g - g\otimes g\otimes
g.B_{a'}\;,\cr=&\,0.\cr}\eqno(27) $$
Hence we see that the Jacobi identity is satisfied if both $a,a'$ are
equivalently normalised solutions of the MYBE, so that
$B_a=B_{a'}={1\over 2}f^d_{bc}t_b\otimes t_c\otimes t_d$ for instance,
or if they are both solutions of the CYBE, $B_a=B_{a'}= 0$.  We can
solve (26) by putting $s=-s'$ and requiring $s$ to be ad-invariant, in
which case, by previous arguments, we see that this combines with (27)
to make (25) a Poisson bracket if $r$ and $r'$ are solutions of the
MYBE.

Therefore, given that (25) is a Poisson bracket on $G$ when $r$ and
$r'$ are solutions of the MYBE with equal but opposite ad-invariant
symmetric parts, we can investigate under what conditions it satisfies
the multiplicative property (23).  Comparing
$$\eqalign{\{f_1,f_2\}(gh)&=<d^Lf_1(gh)\otimes d^Lf_2(gh), r > +
<d^Rf_1(gh)\otimes d^Rf_2(gh), r'>\cr
&= <d(f_1\circ L_g)(h)\otimes d(f_2\circ L_g)(h), L_{h*}r> \cr &\hskip
80pt + <d(f_1\circ R_h)(g)\otimes d(f_2\circ R_h)(g),
R_{g*}r'>\cr}$$
with the sum of the two terms
$$\eqalign{\{f_1\circ L_g, f_2\circ L_g\}(h)&=\, <d(f_1\circ
L_g)(h)\otimes d(f_2\circ L_g)(h),  L_{h*}r + R_{h*}r'>\cr
\{f_1\circ R_h, f_2\circ R_h\}(g)&=\, <d(f_1\circ R_h)(g)\otimes
d(f_2\circ R_h)(g), L_{g*}r + R_{g*}r'>\;,\cr}$$
we see we have Poisson morphisms
$$\eqalign{G_{(r,r')}\times G_{(r,-r)}&\to\,G_{(r,r')}\cr
G_{(-r',r')}\times G_{(r,r')}&\to\,G_{(r,r')}\;.\cr}$$
Here $G_{(r,r')}$ means $G$ taken with the Poisson structure (25), so
that the first Poisson morphism reads
$\{f_1,f_2\}_{(r,r')}(gh)=\{f_1\circ R_h, f_2\circ R_h\}_{(r,r')}(g) +
\{f_1\circ L_g, f_2\circ L_g\}_{(r,-r)}(h)$.
\vskip 3pt
In particular, with $r'=-r$ we have $$G_{(r,-r)}\times
G_{(r,-r)}\to\,G_{(r,-r)}\;,$$ so that $$\{f_1,f_2\} = <d^Lf_1\otimes
d^Lf_2 - d^Rf_1\otimes d^Rf_2, r>\eqno(28)$$ is a Poisson Lie
structure on the group $G$.
\vskip 3pt
There is a direct correspondence between Poisson Lie groups and Lie
bialgebras.  The vector space ${\cal G}^*$ has a Lie bracket defined
on it via $$[\xi_1,\xi_2]_*= d\{f_1,f_2\}(e)\equiv
d_e\{f_1,f_2\}\eqno(29)$$ where $\xi_i=d_ef_i\in {\cal G}^*$.

If we denote by $\Lambda_r$ the Poisson bivector corresponding to the
Poisson Lie structure of $G_{(r,-r)}$, then by defining
$$\displaylines{\eta\,\colon\,G\to\,{\cal G}\wedge {\cal
G}\;,\cr\eta(g)=R_{{g^{-1}}*}\Lambda_r = {\rm
Ad}_g\otimes {\rm Ad}_g.r
- r\;,\cr}$$  we can write the Poisson bracket on $G$ in the form
$$\{f_1,f_2\}=<d^Rf_1\otimes d^Rf_2, \eta>\;,$$
and the Lie bracket on ${\cal G}^*$ as
$$[\xi_1,\xi_2]_*= \eta^*(\xi_1\otimes \xi_2)\;,$$
where $\eta^*$ is the transpose of the tangent map associated with
$\eta$.

Also $$\eta(gh)=R_{{(gh)^{-1}}*}\Lambda_r={\rm Ad}_g\otimes {\rm
Ad}_g.\eta(h) + \eta(g)\;,$$ so that $\eta$ is a 1-cocycle on $G$ with
respect to the adjoint representation of $G$, and
$$\eqalign{<[\xi_1,\xi_2]_*, X>&= <\eta^*(\xi_1\otimes \xi_2),
X>\;,\cr &=<\xi_1\otimes \xi_2,\eta_*(X)>\;,\cr &=<\xi_1\otimes
\xi_2, [X\otimes 1 + 1\otimes X, r]>\;.\cr}$$
This is exactly the Lie bialgebra structure given in (2), here
obtained by differentiating the 1-cocycle on $G$ which defines its
multiplicative Poisson structure.  Hence, (${\cal G},\,{\cal G^*}$) is
referred to as the tangent Lie bialgebra of $G$.
\vskip 15pt
{\biggish VI. Poisson Lie Structure of the Double Group}
\vskip 5pt
We now have all the background material necessary to discuss the
multiplicative Poisson structure of the double group.  We
provide an analysis which, in a similar vein to section III, treats
on an equal footing the construction of the double group from
either a complex semisimple Lie group or its real maximal subgroup.
Previous treatments have not given an explicit description of the
latter case.

For ease of presentation we write $${\cal D = G + G}^*\;$$ where
${\cal G}$ is considered to be {\it either} a complex semisimple Lie
algebra {\it or} its real maximal compact subalgebra.  That is, here
${\cal G}$ denotes either the Lie algebra ${\cal G}$ { or} the Lie
algebra ${\cal K}$ of section III.  Hence the r-matrices $r_\pm\in
{\cal G}\otimes{\cal G}$ are {\it not necessarily} quasitriangular and
the Lie bialgebra (${\cal G^*},\,{\cal G}$) is {\it not necessarily}
factorisable.

 From (28) we have that
the Poisson Lie structure is constructed from an r-matrix, and this
r-matrix was
explicitly
given for the double in (19).  Therefore, the Poisson Lie
structure of the double group can be written
$$\{f_1, f_2\} = <d^Lf_1\otimes d^Lf_2-d^Rf_1\otimes d^Rf_2 ,
(0,t^a)\otimes (t_a, 0)>\eqno(30)$$
for $f_i\in C^\infty(D)$.

By construction, the double Lie algebra leads to a factorisable Lie
bialgebra (${\cal D},\, {\cal D}$), where we identify ${\cal
D}^*={\cal D}$.  So, by (13), we have a unique pointwise decomposition
of the associated group $$\displaylines{\hfill D\sim G\times
G^*\;,\hfill (31)\cr D\ni x=g.\Omega \;,\qquad g\in G\;,\quad
\Omega\in G^*\;,\cr}$$ in a region around the identity.  This
factorisation, because of its particular ordering, leads to what is
known as a dressing action of $G^*$ ($G$) on $G$ ($G^*$).  That is, we
naturally reorder $$\hskip 30pt\Omega.g=g^\Omega.\Omega^g\;,\hskip
40pt \Omega,\Omega^g\in G^*
\;,\quad g,g^\Omega\in G\;,$$
$ g^\Omega$ dependent upon $\Omega$, and $\Omega^g$ dependent upon
$g$.  Fixing $g$ and varying $\Omega$, and vice versa, we obtain the
dressing transformations $$\eqalign{G^*\times G\to\,G\;,\hskip 37pt
g&\mapsto g^\Omega\;,\cr G^*\times G\to\,G^*\;,\hskip 30pt
\Omega&\mapsto
\Omega^g\;.\cr}$$
The importance of dressing transformations is that they are Poisson
actions [3].
\vskip 3pt
The subgroups $G\subset D$ and $G^*\subset D$ are
Poisson Lie subgroups.
It is natural to write the Poisson
structure (30) in terms of functions of  these subgroups.
To do this we note
that $L_{g*}(0,t^a)=(0,L_{g*}t^a)$ and $R_{\Omega *}(t_a,0)=(R_{\Omega
*}t_a,0)$, because they agree with the ordering given in (31).
However,
$$R_{g*}(0,t^a) \hskip 20pt {\rm and } \hskip 20pt L_{\Omega  *}(t_a,
0)$$
are infinitesimal dressing transformations which need to be
calculated.  In [9] a method was briefly indicated when  the bialgebra
(${\cal G},\,{\cal G^*}$) is factorisable, and we give details of this
approach in an appendix.  However, we display here an alternative
calculation which does not use this assumption.

We take a direct approach in which we iteratively reorder the product
and then calculate the commutators produced using the mixed commutator
of the double Lie algebra, relation (18).
Finally, we rewrite the structure constants of ${\cal G}^*$ in terms
of the structure constants of ${\cal G}$ and the components of $r$,
using (6).
$$\eqalign{R_{g*}(0,t^a)\!&\sim t^ag = t^ae^X = t^a(1 + X +
{\textstyle 1\over 2!}X^2 + \ldots) \;,\cr \!&= (1 + X +{\textstyle
1\over 2!}X^2 + \ldots)t^a +(1 + X + \ldots) [t^a, X] + {\textstyle
1\over 2!}\left[[t^a,X],X\right] + \ldots\;,\cr\!  &=e^X\Bigl( t^a
\!+\! X^b(f^a_{bc}t^c-C^{ac}_bt_c) \!+\! {\textstyle 1\over 2!}
X^bX^d\bigl(f^a_{bc}(f^c_{de}t^e
-C^{ce}_dt_e)-C^{ac}_bf^e_{cd}t_e\bigr)\!+\!
\dots \Bigr),\cr
&=e^X\Bigl( t^a + X^bf^a_{bc}t^c + {\textstyle 1\over2!} X^bX^d
f^a_{bc}f^c_{de}t^e + \dots\cr &\qquad\quad - X^b(f^c_{bd}r^{ad} +
f^a_{bd}r^{dc})t_c + {\textstyle 1\over2!} X^bX^d
(f^c_{bf}f^e_{dc}r^{af} - f^a_{bc}f^c_{df} r^{fe})t_e + \dots
\Bigr).\cr}\eqno(32)$$ Introducing the adjoint matrix $L(g)^{ab}\equiv
L^{ab}$ defined by $L_{g*}R_{g^{-1}*}t_a=gt_ag^{-1}= L^{ba}t_b$, it is
straightforward to write (32) in the closed form
$$\eqalign{R_{g*}(0,t^a)&=L_{g*}\bigl(t_b(r^{bc}L^{ac} -
r^{ca}L^{cb}),\; L^{ab}t^b\bigr)\;,\cr &= \bigl(
L_{g*}t_b(r^{bc}L^{ac} - r^{ca}L^{cb}),\;
L^{ab}L_{g*}t^b\bigr)\;.\cr}\eqno(33)$$
The second transformation is manipulated similarly,
$$\eqalign{L_{\Omega *}(t_a, 0)&\sim \Omega t_a= e^{\xi}t_a = (1 +
\xi + {\textstyle 1\over 2!}{\xi}^2  + \ldots)t_a\;,\cr &= t_a(1 + \xi +
{\textstyle 1\over 2!}{\xi}^2 + \ldots) + [\xi,t_a](1 + \xi + \ldots)
+ {\textstyle 1\over 2!}\left[\xi,[\xi,t_a]\right] + \ldots\;,\cr&=
(\chi^{-1})^{ab}R_{\Omega *}t_b + \xi_bf^b_{ac}t^c (1 + \xi + \ldots)
+ {\textstyle 1\over 2!}\xi_b\xi_c(f^b_{ad}C^{cd}_e-
f^c_{de}C_a^{bd})t^e +\ldots \;,\cr}\eqno(34)$$ where
$\chi(\Omega)^{ab}\equiv \chi^{ab}$ is the adjoint matrix for $G^*$.
However, to put this expression into closed form we refer back to the
isomorphism $\tilde I$ of enveloping algebras that we introduced in
section III.  Using its defining relations (16) we can write
$$f^b_{ad}C^{cd}_e-f^c_{de}C_a^{bd}=(ad)_{ae}\circ\tilde
I(t^bt^c)\;,\eqno(35)$$
where ${\rm ad}(t_a)_{bc}=f^b_{ac}$ signifies the adjoint
representation of ${\cal G}$.  Hence, we see that (34) can be written
$$\eqalign{L_{\Omega *}(t_a, 0)&=\bigl( (\chi^{-1})^{ab}R_{\Omega
*}t_b,\; L^{ab}\circ\tilde I(\Omega^{-1})R_{\Omega *}t^b - R_{\Omega
*}t^a\bigr)\;,\cr &= \bigl( (\chi^{-1})^{ab}R_{\Omega *}t_b,\;
iL^{ab}\circ(\tilde r_+-\tilde r_-)(\Omega^{-1})R_{\Omega
*}t^b\bigr)\;,\cr &=\bigl( (\chi^{-1})^{ab}R_{\Omega *}t_b,\;
i[L(\Omega_+^{-1})^{ab} - L(\Omega_-^{-1})^{ab}] R_{\Omega
*}t^b\bigr)\;.\cr}\eqno(36)$$ Here $\tilde r_\pm$ are the extensions
as algebra maps of the r-matrices of the complex semisimple Lie
algebra, as defined in (11).  Hence, we see the appearance of the
r-matrices of the complex semisimple Lie algebra even when we start
out with the real compact subalgebra ${\cal K}$ in the construction of
the double.
\vskip 3pt
We can now write the Poisson structure on the double in terms of
functions of its subgroups $G$ and $G^*$.  For functions of $g\in
G\subset D$ we have $$\eqalign{\{g\otimes g\}(g.\Omega)&= \{g\otimes
g\}(g)\cr &= <(dg,0)\otimes (dg,0), L_{g*}(0,t^a)\otimes
L_{g*}(t_a,0)-R_{g*}(0,t^a)\otimes R_{g*}(t_a,0)>\;,\cr
&=<(dg,0)\otimes (dg,0),(0,L_{g*}t^a)\otimes (L_{g*}t_a,0)> \cr
&\qquad -<(dg,0)\otimes (dg,0),( L_{g*}t_b\{r^{bc}L^{ac} -
r^{ca}L^{cb}\}, L^{ab}L_{g*}t^b)\otimes (R_{g*}t_a,0)>\;,\cr
&=<dg\otimes dg,(r^{ab}-L^{ac}L^{bd}r^{cd})R_{g*}t_a\otimes
R_{g*}t_b>\;,\cr &=[r_{\pm}, g\otimes g]\;,\cr }\eqno(37)$$ where we
use $L^{ba}=L(g^{-1})^{ab}$ since $G$ is semisimple, and we make note
of (3) to see that the result is valid for $r_+$ and $r_-$.  This
well-known result is just the structure of $G_{(r,-r)}$, up to a sign
and is called the Sklyanin bracket [13] on $G$.
\vskip 3pt
For functions of $\Omega\in G^*\subset D$ we find
$$\eqalign{\{\Omega\otimes\Omega\}(g.\Omega)&=
\{\Omega\otimes\Omega\}(\Omega)\;, \cr &= i\bigl(L(\Omega_+^{-1})^{ab}
- L(\Omega_-^{-1})^{ab}\bigr)(\Omega t^a\otimes
t^b\Omega)\;.\cr}\eqno(38)$$ To verify that this is indeed a Poisson
Lie structure on $G^*$ note that (35) can be written
$$\eqalign{(ad)_{ae}\circ\tilde
I(t^bt^c)&=ad_*(t^b)_{ad}\;ad(t_c)_{de}
-ad(t_b)_{ad}\;ad_*(t^c)_{ed}\;,\cr &= ad_*(t^b)_{ad}\; ad_{de}\circ
\tilde I( t^c) - ad_{ad}\circ \tilde I(t^b)\; ad_*(t^c)_{ed}\;,\cr}$$
where $ad_*(t^a)_{bc}= C^{ac}_b$ signifies the adjoint representation
of ${\cal G^*}$.  It follows directly that $$L^{ab}\circ\tilde
I(\Omega'^{-1}\Omega^{-1}) - \delta^{ab} =
\chi(\Omega'^{-1})^{ac} \bigl(
L^{cb} \circ\tilde I(\Omega^{-1}) - \delta^{cb}\bigr) -
\chi(\Omega)^{bc}
\bigl( L^{ac} \circ\tilde
I(\Omega'^{-1})-\delta^{ac} \bigr).$$ This allows us to write
$$\{\Omega\Omega'\otimes\Omega\Omega'\} = \{\Omega\otimes\Omega\}
(\Omega'\otimes\Omega') + (\Omega\otimes\Omega)
\{\Omega'\otimes\Omega'\} $$
which is exactly the multiplicative property (23), as required. As we
have seen in section V, this reflects that
$$\displaylines{\eta_{G^*}\,\colon\,G^*\to\,{\cal G}^*\wedge {\cal
G}^*\cr
\eta_{G^*}(\Omega)= \chi^{ac}\bigl(L(\Omega^{-1}_+)^{cb} -
L(\Omega^{-1}_- )^{cb}\bigr)t^a\otimes t^b}$$ is a 1-cocycle on $G^*$.
\vskip 3pt
The mixed bracket gives
$$\eqalign{\{g\otimes\Omega\}(g.\Omega)=&\;<(dg,0)(g.\Omega)\otimes
(0,d\Omega)(\Omega),\,(0,L_{g.\Omega *}t^a)\otimes L_{\Omega
*}(t_a,0)>\cr &\; - <(dg,0)(g)\otimes
(0,d\Omega)(g.\Omega),\,R_{g*}(0,t^a)\otimes (R_{g.\Omega
*}t_a,0)>\;,\cr =&\;0.\cr}\eqno(39)$$ If we recall (29) we know that
the Lie brackets on ${\cal D}^*$ are obtained by differentiating the
Poisson Lie structure on $D$, and by (20) we require ${\cal D}^*$ to
be a direct sum of Lie algebras, so naturally the mixed bracket must
vanish.  It is straightforward to differentiate the brackets (37) and
(38) at the identity, and to confirm that they lead to ${\cal D}^* =
{\cal G}^*_{opp}\oplus {\cal G}$.
\vskip 3pt
In the case that (${\cal G},\,{\cal G}^*$) is a factorisable Lie
bialgebra it is straightforward to arrive at the usual presentation of
results.  For our purposes it is sufficient to consider
$$\displaylines{ {\cal D= G+G}_R\;, \qquad\qquad D\sim G\times
G_R\;, \cr D\ni x=g.\Omega\;, \qquad\qquad g\in G\;,
\quad \Omega=\Omega_+\Omega_-^{-1}\in G_R\;.\cr}$$
Since we are using factorised group elements the only difference from
above is that we should use the appropriate product in the enveloping
algebra, given by (12) and (14). Hence, we have directly from
(38)
$$\eqalign{\{\Omega\otimes\Omega\}(g.\Omega) &= \{\Omega \otimes
\Omega\} (\Omega)\;,\cr &= i\bigl(L(\Omega_+^{-1})^{ab} -
L(\Omega_-^{-1})^{ab}\bigr)(\Omega *t_a\otimes t_b*\Omega)\;,\cr
&= i\bigl(L(\Omega_+^{-1})^{ab} -
L(\Omega_-^{-1})^{ab}\bigr)\bigl(\Omega_+t_a\Omega_-^{-1}\otimes
(r_+[t^a]\Omega - \Omega r_-[t^a])\bigr)\;,\cr &=
i(r_+.\Omega\otimes\Omega +\Omega\otimes\Omega.r_- -\Omega\otimes
1.r_+.1\otimes\Omega -1\otimes\Omega .r_-.\Omega\otimes 1)\;.\cr
}$$
The final expanded form is how this Poisson bracket is commonly
presented, but it should be remembered that this Poisson bracket is
multiplicative only with respect to the product on $G_R$.
\vskip 15pt
{\biggish VII. The Double of SU(2)}
\vskip 5pt
In order to illustrate the constructions described previously we work
through the example of the double of SU(2).  The Poisson Lie structure
is given in terms of coordinates on SU(2) and SU(2$)^*$, and the
symplectic leaves for these groups are displayed.  It is confirmed
that linearisation of the Poisson brackets leads to the correct Lie
bialgebra structure.
\vskip 3pt
The Pauli matrices provide a basis for su(2) in the fundamental
representation.
Normalised according to our convention in section III they are
$$t_a={i\over  2}\sigma_a\;,\quad t_1={1\over
2}\pmatrix {0&i\cr i&0\cr},\quad t_2 = {1\over  2}\pmatrix {0&1\cr
-1&0\cr},\quad t_3 = {1\over  2}\pmatrix{i&0\cr 0&-i\cr}.$$
The structure constants are
$$[t_a, t_b]=f^c_{ab}t_c\;,\qquad f^c_{ab}=-\epsilon_{abc}\;, $$
where $\epsilon_{abc}$ is the totally antisymmetric Levi-Civita
tensor, and the inner product is given by
$$(t_a,t_b)=-2\; {\rm Trace}(t_at_b)\;.$$
The r-matrix $r_+=r^{ab}t_a\otimes t_b$ , denoted $k_+$ in section
III, has components
$$r^{ab}={1\over 2}\pmatrix{1& 1& 0\cr -1& 1&
0\cr 0 & 0& 1\cr}\;.\eqno(40)$$
This allows us to specify the structure constants of the dual Lie
algebra su(2$)^*$ as
$$C^{ab}_c = f^b_{cd}r^{ad} + f^a_{cd}r^{db}\;,\qquad C^{13}_1 = -
{1\over 2} = C^{23}_2.\eqno(41)$$
A basis for su(2$)^*$ is then
$$t^1={1\over  2}\pmatrix{0 & 1\cr 0 & 0},\quad t^2={1\over
 2}\pmatrix{0 & -i\cr 0 & 0},\quad t^3={1\over
4}\pmatrix{ 1 & 0\cr 0& -1},\eqno(42)$$
with the pairing between the dual vector spaces being given by
$$<t^a,t_b>= 4\;{\rm Im}\;{\rm Trace}
(t^at_b)\;. $$
A group element of $D$, the double of SU(2), is the product of group
elements of
SU(2) and SU(2$)^*$,
$$\eqalign{g=\pmatrix{a & b\cr -\bar b & \bar a}&\in {\rm SU}(2)\;,
\qquad a\bar a + b\bar b = 1\;,\cr \Omega =\pmatrix{u & v + iw\cr 0 &
u^{-1}} &\in {\rm SU}(2)^*\;,\qquad u,v,w\in {\Bbb R}\;, \cr x=
g\Omega =\pmatrix{ au & a(v + iw) +
bu^{-1} \cr -\bar bu & \bar au^{-1} - \bar b(v + iw)} &\in D\simeq
{\rm SL(2, \Bbb C)}\;,\cr}\eqno(43)$$
where we make the natural identification of $D$ with SL(2, ${\Bbb C}$).
\vskip 3pt
The Poisson Lie structure on SU(2), given by
$$\{f_1, f_2\}(g) = < d^Lf_1\otimes d^Lf_2 -d^Rf_1\otimes d^Rf_2,
r_+>\;,$$
reduces to
$$\{a,\bar a\}=-{\textstyle i\over 2}b\bar b\;,\qquad
\{a,b\}={\textstyle i\over 4}ab\;,\qquad \{a, \bar b\}={\textstyle
i\over 4} ac\;,\qquad \{b,\bar b\}=0\;.\eqno(44)$$
In terms of Euler angles we have
$$\displaylines{\hfill g=e^{i\alpha \sigma_3}e^{i\beta
\sigma_2}e^{i\gamma \sigma_3}\in {\rm SU}(2)\;, \hfill (45)\cr
\{\alpha, \sin\beta\}={\textstyle 1\over 8} \sin\beta\;,\qquad
\{\alpha, \gamma \}=0\;,\qquad \{\gamma , \sin\beta\}={\textstyle
1\over 8} \sin\beta\;.\cr}$$
To verify that these relations generate the structure of su(2$)^*$
when linearised, note that the basis elements of su(2$)^*$ follow from
$$g=e^{\lambda^at_a}\in {\rm SU}(2)\;,\qquad d_e\lambda^a=t^a \in
su(2)^*
\eqno(46)$$
We have $$\tilde \lambda^1=\sin(\alpha - \gamma)\sin \beta\;,\qquad
\tilde
\lambda^2 = \cos(\alpha - \gamma)\sin \beta\;,\qquad \tilde \lambda^3
= \sin(\alpha + \gamma)\cos\beta\;,$$ where $\tilde
\lambda^i=(\lambda^i/\theta)\sin(\theta/2)$, and $\theta^{\,2} =
\sum_i (\lambda^i)^2$.  For these variables the Poisson brackets are
$$\{\tilde \lambda^1,\tilde \lambda^2\}=0\;,\qquad\{\tilde
\lambda^1,\tilde \lambda^3 \}=-{\textstyle 1\over 4}\cos(\theta/2) \tilde
\lambda^1 \;,\qquad \{\tilde \lambda^2,\tilde \lambda^3 \}=-{\textstyle
1\over 4}\cos(\theta /2)\tilde \lambda^2\;.$$ If we implement (29),
that is, linearise the Poisson structure, we obtain $$[d_e\lambda^1,
d_e\lambda^2]_*=0\;,\qquad [d_e\lambda^1, d_e\lambda^3]_*=
-{\textstyle 1\over 2}d_e\lambda^1\;,\qquad [d_e\lambda^2,
d_e\lambda^3]_*= -{\textstyle 1\over 2}d_e\lambda^2 \;,
\eqno(47) $$
where we have used $d_e\lambda^i=2d_e\tilde\lambda^i$.  These brackets
give rise to the structure constants in (41), as required.
\vskip 3pt
We remark that it is straightforward to display the symplectic leaves
of this Poisson structure on SU(2).  It is clear that $\alpha -
\gamma$ is a
Casimir, so the symplectic leaves are specified by $\alpha - \gamma =
c$, a constant.  Hence we have an $S^1$ of two-dimensional leaves
given by
$$g=e^{i\alpha \sigma_3}e^{i\beta \sigma_2}e^{i\alpha \sigma_3} e^{ic
\sigma_3}\sim\pmatrix{e^{i\alpha}\cos\beta & e^{ic}\sin\beta \cr -
e^{-ic} \sin\beta & e^{-i\alpha}\cos\beta\cr},\eqno(48)$$
except for $\beta = 0$, for which we have an $S^1$ of zero-dimensional
leaves
$$g=e^{ic\sigma_3}=\pmatrix{e^{ic} & 0\cr 0 & e^{-ic}}.\eqno(49)$$
More general work on the symplectic leaves of compact Poisson Lie
groups can be found in [14] and in references therein.
\vskip 3pt
The Poisson structure on SU(2$)^*$ is given by
$$\{f_1, f_2\}(\Omega) = < d^Rf_1\otimes d^Rf_2, i\chi^{ac}\bigl(
L(\Omega_+^{-1})^{cb} - L(\Omega_-^{-1})^{cb}\bigr)t^a\otimes t^b>\;.
$$
Since, by (43), $\Omega$ is dependent only on two variables, one real
and one complex, we need to consider $$\{\Omega\otimes
\bar\Omega\}=i\chi^{ac}\bigl(L(\Omega_+^{-1})^{cb} -
L(\Omega_-^{-1})^{cb}\bigr) t^a\Omega\otimes \bar
t^b\bar\Omega\;,\eqno(50)$$ where a bar indicates complex conjugation,
in order to calculate all the relations.

As before, the basis elements of the dual Lie algebra, in this case
su(2), come from
$$\Omega= e^{\mu_at^a} = \pmatrix{ e^{\mu_3/4} & 2{\textstyle {(\mu_1 -
i\mu_2)\over \mu_3}}\sinh \mu_3/4\cr 0 & e^{-\mu_3/4}},\qquad d_e\mu_a
= t_a \in su(2)\;.\eqno(51)$$
We find
$$\Omega_+= \pmatrix{e^{\mu_3/4} & 2{\textstyle
{\mu_-\over \mu_3}}\sinh \mu_3/4\cr 0 & e^{-\mu_3/4}},\qquad \Omega_-
= \pmatrix{e^{-\mu_3/4} & 0\cr -2{\textstyle {\mu_+\over \mu_3}}\sinh
\mu_3/4 & e^{\mu_3/4}}, $$
where $\Omega_+,\Omega_- \in {\rm SL}(2,{\Bbb C})$, and consequently
$$\eqalign{\chi^{1c}\bigl(L(\Omega_+^{-1})^{c2} -
L(\Omega_-^{-1})^{c2}\bigr)& =  2ie^{\mu_3/2}(\sinh\mu_3/2 + 2
{\textstyle {\mu_+\mu_-\over \mu_3}}\sinh^2 \mu_3/4)\;,\cr
\chi^{1c}\bigl(L(\Omega_+^{-1})^{c3} -L(\Omega_-^{-1})^{c3}\bigr) & =
-4ie^{\mu_3/4} {\textstyle {\mu_2 \over \mu_3}}\sinh \mu_3/4 \;,\cr
\chi^{2c}\bigl(L(\Omega_+^{-1})^{c3} -L(\Omega_-^{-1})^{c3}\bigr) & =
4ie^{\mu_3/4} {\textstyle {\mu_1 \over \mu_3}}\sinh \mu_3/4\;,\cr } $$
where $\mu_\pm = \mu_1\pm i\mu_2$.  It follows that (50) gives rise
to the relations
$$\displaylines{\hfill \{\mu_3, \mu_+\} = i \mu_+\;,\qquad \{\mu_3,
\mu_-\} = -i \mu_-\;,\hfill (52)\cr
\{\mu_+, \mu_-\} = {\textstyle i\over 2}\mu^2_3 \coth\mu_3/4 +
2i\mu_+\mu_-\left( {\textstyle 1\over 4} \coth\mu_3/4 - {1\over \mu_3}
\right)\;.\cr}$$
Linearisation produces
$$[d_e\mu_3,d_e\mu_+] = i d_e\mu_+\;,\qquad [d_e\mu_3,d_e\mu_-] = i
d_e\mu_- \;,\qquad [d_e\mu_+, d_e\mu_-]= 2id_etex\mu_3\;,$$
which is the structure of su(2) written in terms of the generators
$t_1 \pm it_2,\, t_3$.
\vskip 3pt
It is not difficult to find that ${1\over 2}\cosh(\mu_3/2) +
(\mu_+\mu_-/ \mu_3^2)\sinh^2(\mu_3/4)$ is a Casimir for the Poisson
structure of SU(2$)^*$.  Therefore,  we have two-dimensional symplectic
leaves given by
$$\Omega=\pmatrix{ e^x & ze^{iy}\cr 0 &
e^{-x}},\qquad e^{2x} + e^{-2x} + z^2= c^2 \eqno(53)$$
where $x,\,y,\,z\in \Bbb R$,  and $c$ is a real constant, and the
zero-dimensional leaf
$$\Omega=\pmatrix{ 1 & 0\cr 0 & 1}. \eqno(54)$$
\vskip 15pt
{\biggish VIII. Concluding Remark}
\vskip 5pt
We have given a detailed  account of the Poisson Lie
structure of the double of a semisimple group, developing all the basic
elements required.  This structure has been displayed completely
explicitly and its usage made straightforward.
\vskip 3pt
Closely associated to the Poisson Lie structure of the double is
a second  Poisson  structure given by
$$\{f_1, f_2\} = <d^Lf_1\otimes d^Lf_2, \bar r> - <d^Rf_1\otimes d^Rf_2,
\tau (\bar r)>$$
For the double as developed in section VI this leads to the relations
$$\eqalign{\{g\otimes g\} &= [g\otimes g, r_\pm]\;,\cr
\{\Omega\otimes \Omega\} &= i\bigl(L(\Omega_+^{-1})^{ab}
- L(\Omega_-^{-1})^{ab}\bigr)(\Omega t^a\otimes t^b\Omega)\;.\cr
\{g\otimes \Omega\} &= -gt_a\otimes t^a\Omega\;.\cr}$$
This structure is symplectic for the double group considered to be
connected and simply connected [2].  It should be interesting to
extend our analysis to give an algebraic description of this symplectic
structure.  A full geometric treatment has been given by Alekseev and
Malkin [15].
\vskip 15pt
{\biggish Acknowledgements}
\vskip 5pt
I should like to thank Alan Macfarlane for his guidance and his
encouragement, and also Stuart Rankin and Shahn Majid for
helpful discussions.  This work was supported by a S.E.R.C. research
studentship.
\vskip 15pt
{\biggish Appendix - Alternative Calculation of Dressing Transformations}
\vskip 5pt
We give here an explanation of how the infinitesimal dressing
transformations calculated in section VI may be obtained by an
alternative method, applicable in the particular case of the double
constructed from a factorisable
Lie bialgebra.  We elaborate on brief details given in [9].
\vskip 3pt
For a factorisable Lie bialgebra (${\cal G},\,{\cal G}^*$) it was
shown in section IV how the double can be considered in the form
$$\eqalign{{\cal D = G + G^*} &\to {\cal D' = G\oplus G = G + G}_R\;,\cr
{\cal D \supset G} \ni (t_a,0) &\mapsto (t_a, t_a)\in {\cal G \subset
D'}\;, \cr {\cal D \supset G}^* \ni (0, t^a) &\mapsto (r_+t^a,
r_-t^a)\equiv (t_a^+, t_a^-)
\in {\cal G}_R \subset {\cal D'} \;.\cr}$$
For convenience we consider here $r_+ -r_- = I$ which differs by an
unimportant
factor of $i$ to the convention used in sections III and  VI.

The corresponding double group is $$\eqalign{ D'&=G\times G\sim
G\times G_R\;,
\cr D'&\ni x=(g,g).(\Omega_+,\Omega_-)\;,\qquad\qquad
g\in G\;,\quad \Omega=\Omega_+\Omega_-^{-1}\in G_R\;.\cr}\eqno({\rm
A}.1)$$ The infinitesimal dressing transformations we want to
calculate become $$\eqalign{R_{g*}(0,t^a)&\mapsto R_{(g,g)*}(t_a^+,
t_a^-) = L_{(g,g)*} {\rm Ad}_{(g^{-1},g^{-1})} (t_a^+, t_a^-)\;,\cr
L_{\Omega *}(t_a,0)&
\mapsto L_{(\Omega_+,\Omega_-)*}(t_a, t_a) = R_{(\Omega_+,\Omega_-)*}
{\rm Ad}_{(\Omega_+,\Omega_-)} (t_a, t_a)\;,\cr}\eqno({\rm A}.2)$$
where Ad signifies the adjoint representation of the double group.

We have $$\eqalign{(L(g)^{ba}t_a, 0)&\mapsto {\rm Ad}_{(g,g)}(t_a,
t_a) = (gt_ag^{-1},\; gt_ag^{-1}) \;,\cr (0,
X(\Omega)^{ba}t^a)&\mapsto {\rm Ad}_{(\Omega_+,\Omega_-)}(t_a^+,
t_a^-) = (\Omega_+ t_a^+\Omega_+^{-1},
\;\Omega_- t_a^-\Omega_-^{-1} )\;,\cr}\eqno({\rm A}.3)$$
and also $$\eqalign{{\rm Ad}_{(g^{-1},g^{-1})} (t_a^+, t_a^-)&=
(g^{-1}t_a^+g,\; g^{-1}t_a^-g )\;,\cr {\rm
Ad}_{(\Omega_+,\Omega_-)}(t_a, t_a)& = (\Omega_+
t_a\Omega_+^{-1},\;\Omega_- t_a\Omega_-^{-1} ) \;.\cr}\eqno({\rm
A}.4)$$ These need to be rewritten in terms of elements of ${\cal
G},\, {\cal G}_R \subset {\cal D'} $ and this can be done essentially
by observation.

We see that
$${\rm Ad}_{(g,g)}(t_a, t_a) = \bigl( g^{-1}t_a^+g - (g^{-1}t_ag)^+,\,
g^{-1}t_a^-g - (g^{-1}t_ag)^-\bigr) + \bigl( (g^{-1}t_ag)^+,\,
(g^{-1}t_ag)^-\bigr) \;,\eqno({\rm A}.5)$$
where
$$g^{-1}t_a^+g - (g^{-1}t_ag)^+ = g^{-1}t_a^-g - (g^{-1}t_ag)^- =
(r^{bc}L^{ac} - r^{ca}L^{cb})t_b\;. $$
Hence,  we deduce that (A.5) is the image of
$${\rm Ad}_g(0,t^a)= \bigl( (r^{bc}L^{ac} - r^{ca}L^{cb})t_b,\;
L(g^{-1})^{ba}t^b \bigr)\eqno({\rm A}.6)$$
which leads directly to (33).

Similarly, we can write
$$\eqalign{{\rm Ad}_{(\Omega_+,\Omega_-)}(t_a, t_a) &= (\Omega_-
t_a^+\Omega_-^{-1} - \Omega_+ t_a^-\Omega_+^{-1},\; \Omega_-
t_a^+\Omega_-^{-1} - \Omega_+ t_a^-\Omega_+^{-1})\cr &
\qquad + (\Omega_+ t_a^+\Omega_+^{-1} - \Omega_- t_a^+\Omega_-^{-1},
\; \Omega_+ t_a^-\Omega_+^{-1}- \Omega_- t_a^-\Omega_-^{-1})\;.\cr}
\eqno({\rm A}.7)$$
The second term can be written
$$(\Omega_+ t_a^+\Omega_+^{-1} - \Omega_- t_a^+\Omega_-^{-1},
\; \Omega_+ t_a^-\Omega_+^{-1}- \Omega_- t_a^-\Omega_-^{-1}) = \bigl(
L(\Omega_+ )^{ba} - L(\Omega_- )^{ba}\bigr)(t_a^+, t_a^-) $$
whilst for the first term of (A.7) we use (A.4) to note  that
$$X^{ba}t_b = \Omega_+ t_a^+\Omega_+^{-1} - \Omega_-
t_a^-\Omega_-^{-1} \Longleftrightarrow X^{ab}= L(\Omega_+)^{ac}r^{bc}
+ L(\Omega_-)^{ac}r^{cb}.$$
It follows that we can write (A.7) as
$${\rm Ad}_{(\Omega_+,\Omega_-)}(t_a, t_a) =
X(\Omega^{-1})^{ab}(t_b,t_b) + \bigl(
L(\Omega_+ )^{ba} - L(\Omega_- )^{ba}\bigr)(t_a^+, t_a^-)\eqno({\rm
A}.8)$$
which is clearly the image of
$${\rm Ad}_\Omega (t_a, 0) = \bigl(X(\Omega^{-1})^{ab}t_b, [L(\Omega_+
)^{ba} - L(\Omega_- )^{ba} ]t^b\bigr)\;.\eqno({\rm A}.9)$$
This leads directly to (36), up to the factor of $i$ mentioned above.
\vskip 15pt
{\biggish References}
\vskip  3pt
\parindent=15pt
\item{[1]} N.Y. Reshetikhin and M.A. Semenov-Tian-Shansky,
 J. Geom. Phys. {\bf 5}
(1988) 533-550.
\item{[2]}  V.G. Drinfeld,
 Soviet Math. Dokl. {\bf 27} (1983) 68-71.
\item{[3]}  M.A. Semenov-Tian-Shansky, RIMS
Kyoto Univ. {\bf 21} (1985)
1237-1260.
\item{[4]} J.H. Lu and A. Weinstein,
J. Diff. Geom. {\bf 31}
(1990) 501-526.
\item{[5]} S. Helgason, {\it Differential Geometry, Lie Groups and
Symmetric Spaces} (Academic Press, New York, 1978).
\item{[6]} S. Majid,
Pacific J. Math. {\bf 141}
(1990) 311-332.
\item{[7]} Y. Kosmann-Schwarzbach and F. Magri,
Ann. Inst. Henri Poincar\'e (Physique\hfill \break
th\'eorique) {\bf 4} (1988) 433-460.
\item{[8]} O. Babelon and D. Bernard,
Phys. Lett. {\bf B260}
(1991) 81-86.
\item{[9]} M.A. Semenov-Tian-Shansky
in {\it Field Theory, Quantum gravity and Strings II}, Lecture Notes
in Physics, {\bf 280} (Springer-Verlag, Berlin, 1987), 174-214.
\item{[10]} P. Goddard and D. Olive,
Int. J. Mod. Phys. {\bf A1}
(1986) 303-414.
\item{[11]} M. Jimbo in {\it Field Theory, Quantum Gravity and
Strings}, Lecture Notes in Physics, {\bf 246} (Springer-Verlag,
Berlin, 1986), 335-361.
\item{[12]} J.H. Lu, Ph.D Thesis, Berkeley (1990).
\item{[13]} E.K. Sklyanin, Funct. Anal. Appl. {\bf 16} (1983) 263-270.
\item{[14]} Y.S. Soibelman, Int. J. Mod. Phys. {\bf A7} (1992) 859-887.
\item{[15]} A.Y. Aleekseev and A.Z. Malkin, preprint PAR-LPTHE
93-08, to appear in Comm. Math. Phys.
\end